\documentclass[12pt,preprint]{aastex}
\usepackage{emulateapj5}
\usepackage{apjfonts}
\newcommand{\Msolar}{\mbox{\,$\rm M_{\odot}$}}        
\newcommand{\Lsolar}{\mbox{\,$\rm L_{\odot}$}}        
\newcommand{\kms}{km~s$^{-1}$}

\newcommand{\ergs}{\rm erg~s^{-1}}
\newcommand{\OIII}{[\rm O~ \sc{III}]}
\newcommand{\LOIII}{L_{\rm [O~ \sc{III}]}}

\begin{document}

\title{The accretion ratios in Seyfert 2 galaxies with and
without hidden broad-line regions}

\author{W. Bian \altaffilmark{1,}\altaffilmark{2} and Q. Gu\altaffilmark{3}} \altaffiltext{1}{Key Laboratory for Particle Astrophysics,
Institute of High Energy Physics, Chinese Academy of Sciences,
Beijing 100039, China} \altaffiltext{2}{Department of Physics and
Institute of Theoretical Physics, Nanjing Normal University,
Nanjing 210097, China} \altaffiltext{3}{Department of Astronomy,
Nanjing University, Nanjing 210093, China}
\email{whbian@njnu.edu.cn}

\shorttitle{The accretion ratios in Sy2s}

\shortauthors{Bian \& Gu}

\slugcomment{Received 2006 July 10; accepted 2006 Oct 30}

\begin{abstract}
Using a large sample of 90 Seyfert 2 galaxies (Sy2s) with
spectropolarimetric observations, we tested the suggestion that
the presence of hidden broad-line regions (HBLRs) in  Sy2s is
dependent upon the Eddington ratio. The stellar velocity
dispersion and the extinction-corrected $\OIII$ luminosity are
used to derive the mass of central super-massive black holes and
the Eddington ratio. We found that: (1) below the Eddington ratio
threshold of $10^{-1.37}$, all but one object belong to non-HBLRs
Sy2s; while at higher Eddington ratio, there is no obvious
discrimination in the Eddington ratio and black hole mass
distributions for Sy2s with and without HBLRs; (2) nearly all
low-luminosity Sy2s (e.g., $\LOIII < 10^{41} \ergs$) do not show
HBLRs regardless of the column density of neutral hydrogen
($N_{\rm H}$); (3) for high-luminosity Sy2s, the possibility to
detect HBLRs Sy2s is almost the same as that of non-HBLRs Sy2s;
(4) when considering only Compton-thin Sy2s with higher $\OIII$
luminosity ($>10^{41} \ergs$), we find a very high detectability
of HBLRs ,$\sim$ 85\%. These results suggested that AGN luminosity
plays a major role in  not detecting HBLRs in low-luminosity Sy2s,
while for high-luminosity Sy2s, the detectability of HBLRs depends
not only upon the AGN activity, but also upon the torus
obscuration.
\end{abstract}

\section{INTRODUCTION}

Seyfert 2 galaxies (Sy2s) belong to a subclass of low-luminosity
active galactic nuclei (AGN) with the absence of broad permitted
optical lines, compared to Seyfert 1 galaxies (Sy1s). The standard
paradigm of AGN is an accretion disk surrounding a central
super-massive black hole (SMBH), with other components, such as
broad-line regions (BLRs), narrow-line regions (NLRs), jet, torus,
et al. (e.g. Rees 1984; Antonucci 1993; Urry \& Padovani 1995). In
the scheme of AGN oriented-unification model, the distinction
between Sy2s and Sy1s depends on whether the central engine and
BLRs are viewed directly (Sy1s) or are obscured by circumnuclear
torus (Sy2s). This scenario was first suggested to explain the
presence of polarized broad emission lines in NGC 1068, in addition to
spectropolarimetric observations of hidden broad-line regions
(HBLRs) in several other Sy2s (Antonucci \& Miller 1985; Tran
1995; Heisler, Lumsden \& Bailey 1997; Moran et al. 2000; Lumsden
et al. 2001; Tran 2001; Zakamsa 2003b). Infrared observations also
confirmed the existence of broad emission lines in several Sy2s
(Veilleux, Goodrich \& Hill 1997). More evidence for this
simple unification model comes from the X-ray observations: Sy2s
show much higher column density of neutral hydrogen ($N_{\rm H}$)
than type 1 ones as expected from the torus obscuration. In the
local Universe, about half of Sy2s are found to be Compton-thick sources with $N_{\rm
H}> 10^{24}$ cm$^{-2}$ (Maiolino et al. 1998; Bassani et al. 1999;
Risaliti, Maiolino \& Salvati 1999).

However, some Sy2s don't show HBLRs in the spectropolarimetric
observations and some Sy2s have column densities lower than
$10^{22} \rm cm^{-2}$ (so called "unabsorbed Sy2s", Panessa \&
Bassini 2002) in the X-ray observation, which challenged the
oriented-unification model indeed. It is still not clear what kind
of physical process is related to the presence of HBLRs in Sy2s.
Natively, the detectability of HBLRs in Sy2s depends upon the
sensitivity of observation, the strength of BLRs, and the
inclination of the torus to the line of sight, etc (e.g. Moran et
al. 2001; Gu \& Huang 2002; Nicastro et al. 2003; Laor 2003; Tran 2003). For
unabsorbed Sy2s with $N_{\rm H}<10^{22} \rm cm^{-2}$, their BLRs
are either obscured by something other than the torus or it is
weak or absent (Panessa \& Bassini 2002; Gallo et al. 2006). For
Compton-thick Sy2s ($N_{\rm H}> 10^{24} \rm cm^{-2}$), it is
impossible to estimate precisely the intrinsic X-ray luminosity
(e.g. Maiolino et al. 1998; Bassani et al. 1999; Shu et al. 2006).

It has been shown that the detectability of HBLRs in Sy2s is found
to be related with the IRAS $f_{60}/f_{25}$ flux ratio (Heisler,
Lumsden \& Bailey 1997; Lumsden, Alexander \& Hough 2004) and the
AGN activity (Sy2s with HBLRs have higher AGN luminosities, e.g.
Lumsden \& Alexander 2001; Gu \& Huang 2002; Tran 2001, 2003).
Some authors claimed that Sy2s without HBLRs have larger $N_{\rm
H}$ (e.g. Lumsden, Alexander \& Hough 2004) and some did not (e.g.
Gu \& Huang 2002; Tran 2001). Recently, Shu et al.(2006) suggested
that the detectability of HBLRs for luminous Sy2s is related to
$N_{\rm H}$. For a small spectropolarimetric sample (from Tran
2001), Nicastro et al.(2003) found that the detectability of HBLRs
in Sy2s is regulated by the Eddington ratio. Laor (2003) also
proposed an analogous model for the existence of BLRs in AGNs.

In this paper, we calculate the Eddington ratios for a larger Sy2s
sample (mainly from Gu \& Huang 2002) to test the suggestion that
the presence of HBLRs in Sy2s is dependent upon the Eddington
ratios. In section 2, we briefly introduce the sample and present
the calculation of Eddington ratios. Our results and discussion
are given in Sec. 3.

\section{The Sample and Eddington Ratios}
Gu \& Huang (2002) collected host-galaxy morphological types and multi-wavelength data
(radio, infrared, optical, and hard X-ray) for a sample of 90 Seyfert 2 galaxies
with spectropolarimetric observations. Out of these 90 objects, 41
show HBLRs, and 49 do not (see their Tables 1 and 2). The
HBLR/non-HBLR classification is mostly from the results of Tran
(2001; 2003).

In order to calculate the Eddington ratios, $L_{\rm bol}/L_{\rm
Edd}$, where $L_{\rm Edd}=1.26\times 10^{38}M_{\rm
BH}/\Msolar~\ergs$, we need to know the SMBH mass and the
bolometric luminosity. As we know, recent important progress on
AGN study is that we can derive more reliable SMBHs' masses
through several empirical methods. The broad emission lines from
BLRs (e.g. H$\beta$, Mg II, CIV, H$\alpha$), the reverberation
mapping method, the empirical BLR size-luminosity relation and the
SMBHs mass-stellar velocity dispersion relation (the $M_{\rm
BH}-\sigma_{*}$ relation) can be used to derive the virial SMBHs
masses in type 1 AGNs (e.g. Kaspi et al. 2000, 2005; Wang \& Lu
2001; Gu et al. 2001; Vestergaard 2002; McLure \& Jarvis 2002;
Bian \& Zhao 2004; Wu et al. 2004; Greene \& Ho 2006a; Nelson et
al. 2001; Tremaine et al. 2002; Greene \& Ho 2006a, 2006b; Bian et
al. 2006).

Because of the absence of the broad emission lines from BLRs in
Sy2s, we use the $M_{\rm BH}-\sigma_{*}$ relation to derive the
SMBHs mass (Tremaine et al. 2002), which is $M_{\rm BH} (\Msolar)=
10^{8.13}[\sigma_{*}/(200 \rm km~ s^{-1})]^{4.02} $. The stellar
velocity dispersions ($\sigma_{*}$) are collected from the recent
literature, mainly from Nelson \& Whittle (1995) and
Garcia-Rissmann et al. (2005), the latter measured $\sigma_{*}$
through fitting Ca II$\lambda \lambda$ 8498, 8542, 8662 triplet.
For the common objects in Nelson \& Whittle (1995) and
Garcia-Rissmann et al. (2005), the values of $\sigma_{*}$ are
almost the same and we adopted the new values by Garcia-Rissmann
et al. (2005) for their smaller error bars. Thus, we obtained
$\sigma_*$ for 20 HBLRs Sy2s and 25 non-HBLRs Sy2s. Because the
NLRs dynamics is primarily dominated by the bulge gravitational
potential (e.g. Whittle 1992; Nelson \& Whittle 1996; Greene \& Ho
2005), we also used full width at half-maximum (FWHM) of $\OIII
\lambda$5007 as a proxy for $\sigma_*$ for Sy2s without direct
$\sigma_*$ measurement, $\sigma_{*}=FWHM_{\OIII}/2.35/1.34$, where
1.34 is a factor between the stellar velocity dispersion and the
gas velocity dispersion (Greene \& Ho 2005). We obtained
$\sigma_*$ from [O III] FWHM for 4 HBLRs Sy2s and 11 non-HBLRs
Sy2s. Finally, we could calculate the SMBHs masses for 24 HBLRs
and 36 non-HBLRs Sy2s.

Following Kauffmann et al. (2003), we use the $\OIII$ luminosity as a
tracer of the AGN intrinsic luminosity (Zakamsa
et al. 2003a; Heckman et al., 2004; Greene \& Ho 2005). The
extinction-corrected luminosity of  $\OIII \lambda$5007 is given as
$\LOIII = 4 \pi D^{2}F_{[\rm O III]}^{\rm cor}$, where F$_{[\rm O
III]}^{\rm cor}$ is the extinction-corrected flux of $\OIII
\lambda$5007 emission line derived from the relation (Bassani et al.
1999)
$$F_{\rm [O III]}^{\rm cor} = F_{\rm [O III]}^{\rm obs} \rm
[\frac{(H_\alpha/H_\beta)_{obs}}{(H_\alpha/H_\beta)_{0}}]^{2.94}
$$

\noindent where an intrinsic Balmer decrement $\rm (H_\alpha/H_\beta)_{0} =
3.0$ is adopted. The bolometric luminosity is calculated by assuming $L_{\rm bol}=3500L_{\rm [O III]}$ (with an
uncertainty of 0.4 dex, Heckman et al., 2004). At last we
calculate the Eddington ratios, $L_{\rm bol}/L_{\rm Edd}$, for
this larger sample of Sy2s.

The uncertainties in the Eddington ratios depend on the
uncertainties in black hole masses and bolometric
luminosities. The uncertainties of SMBHs are from $\sigma_{*}$ and
$M_{\rm BH}-\sigma_{*}$ relation (0.3 dex, Tremaine et al. 2002).
The uncertainty of $\sigma_{*}$ is typically about 20 \kms, which
would lead to uncertainty of about 0.3 dex in the logarithm of
SMBH mass. Combined with the uncertainties of $M_{\rm
BH}-\sigma_{*}$ relation and the bolometric luminosity, the
uncertainties of the Eddington ratio is about 0.5 index. Recently,
Zhang \& Wang (2006) used the $\OIII$ FWHM to calculated the SMBHs
masses of Sy2s and suggested that absorbed non-HBLRs Sy2s harbor
less massive black hole with higher accretion rates, which are similar
to narrow-line Seyfert 1 galaxies (NLS1s). They found that the mean
Eddington ratio of non-HBLRs Sy2s is $0.23 \pm 0.14$, which is
larger than ours ($-0.47 \pm 0.16$). The difference is due to the
SMBHs masses. Their SMBHs masses were derived from the
FWHM$_{\OIII}$ while ours are mainly from the stellar velocity
dispersion. In Figure 1, we showed the relation between the
$\OIII$ FWHM and $\sigma_{*}$ for the common objects. The solid
line showed $\sigma_{*}=FWHM_{\rm [O III]}/2.35/1.34$. It is
obvious that, for most objects in Zhang \& Wang (2006), the
$\OIII$ FWHM underestimates $\sigma_{*}$.

In Table 1 and Table 2, we presented the multi-wavelength properties  of non-HBLRs Sy2s
and HBLRs Sy2s. In Table 3, we presented the distributions of SMBHs
masses and the Eddington ratios for Sy2s with and without HBLRs.

\section{Results and discussion}

\subsection{Distributions of SMBHs masses and the Eddington ratios}

In Figure 2, we show the distributions of the SMBHs masses and
the Eddington ratios for HBLRs Sy2s and non-HBLRs Sy2s. The
distribution of the SMBHs masses is almost the same for HBLRs Sy2s and
non-HBLRs Sy2s. For the whole sample, the mean Eddington ratio for
HBLRs Sy2s is larger than non-HBLRs Sy2s by the magnitude of 0.40,
which is much smaller than the standard deviation of
0.79 (see Table 3).

In order to show the similarity between the distributions for the
Eddington ratios and the SMBHs masses, we used the two sample
Kolmogorov-Smirnov (K-S) test, {\it kolmov} task in IRAF \footnote{IRAF is distributed by the National
 Optical Astronomy Observatories, which are operated by the
 Association of Universities for Research in Astronomy, Inc., under
 cooperative agreement with the National Science Foundation.}.
The K-S test showed that the distributions of the Eddington
ratios/SMBHs masses for non-HBLRs Sy2s and HBLRs Sy2s subsamples
are drawn from the same parent population with the probability of
48.5\% and 64.8\%, respectively (see Table 3 for detail).
Therefore, for the whole sample, there is no significant
difference in the Eddington ratios/SMBHs masses between non-HBLRs
Sy2s and HBLRs Sy2s.

Due to the complicated nature of unabsorbed Sy2s ($N_{\rm H} <
10^{22}\rm cm^{-2}$, Panessa \& Bassini 2002; Gallo et al. 2006 ),
we also calculated the distributions of the SMBHs masses and the
Eddington ratios only for HBLRs Sy2s and non-HBLRs Sy2s with
$N_{\rm H} > 10^{22}\rm cm^{-2}$. The distributions of SMBHs
masses and the Eddington ratios keep nearly the same as for the
whole Sy2s sample.

As we showed in Section 2, the [O III] FWHM
underestimated $\sigma_{*}$ if we used the formulae,
$\sigma_{*}=FWHM_{\rm [O III]}/2.35/1.34$. In our sample, there
are 15 objects without direction $\sigma_*$ measurements.
Excluding these 15 objects, we perform the K-S test to Sy2s with
direct $\sigma_*$ measurements, the results are almost the same,
which is presented in Table 3 as case C.

\subsection{An Eddington ratio threshold for HBLRs and non-HBLRs Sy2s?}

As we mentioned in Section 1, AGN activity is required to
understand the difference of Sy2s types. AGN luminosity is mainly
from the disk accretion onto central SMBHs. Some authors discussed
the relation between the formation of BLRs and the disk accretion
process (e.g. Nicastro 2000; Bian \& Zhao 2002). Nicastro (2000)
suggested that BLRs were formed by accretion disk instabilities
occurring in the critical radius at which the disk changed from
gas pressure dominated to radiation pressure dominated. This
radius diminishes with decreasing low enough accretion rates and
BLRs can't form. Laor (2003) also proposed an analogous model that
the existence of BLRs in AGNs was based on the observed upper limit
of emission line width of 25,000 \kms. Nicastro et al. (2003) used
the relation between the SMBHs masses and the Hubble-type-corrected
bulge luminosity to calculate the SMBHs masses. They
then derived the Eddington ratio from the 2-10 keV X-ray
luminosity ($L_{\rm X}$) by assuming a conversion factor,
$L_{\rm bol}/L_{\rm X}$, of 10. They
argued that Sy2s with HBLRs have the Eddington ratio larger
than the threshold of $10^{-3}$, while non-HBLRs Sy2s lie below
this threshold (see their Figure 2).

Though 2-10keV hard X-ray luminosity is a direct estimator of
the AGN activity, and $ \OIII \lambda$5007 luminosity represents
only an indirect (i.e., reprocessed) measurement of the nuclear
activity, due to the large absorbing column density of Sy2s, it is
very hard and time-consuming to derive X-ray data for a large
sample of Sy2s. On the other side, Kauffmann et al. (2003) have
shown that extinction-corrected $\LOIII$ is a good indicator of
AGN activity for Type II AGNs. Here we obtained more reliable
SMBHs masses from the Tremaine's $M_{\rm BH}-\sigma_{*}$ relation,
and the bolometric luminosity from the extinction-corrected $
\OIII \lambda$5007 luminosity.   We used $L_{\rm bol}=3500 \LOIII$
and Nicastro et al. (2003) used $L_{\rm bol}=10L_{\rm X}$.
However, some larger samples showed that $log(L_{\rm X}/ \LOIII)
\sim$ 1 dex (e.g. Heckman et al. 2005), therefore the $L_{\rm
bol}/L_{\rm Edd}$ from Nicastro et al. (2003) would be lower than
ours by $\sim$ 1.5 dex. In Nicastro et al. (2003), the threshold
of Eddington ratio is corresponding to the $L_{\rm X}/L_{\rm Edd}$
threshold of $10^{-4}$. In order to test the suggestion of
Nicastro et al. (2003), we have to transform the $L_{\rm X}/L_{\rm
Edd}$ threshold into the $L_{\rm bol}/L_{\rm Edd}$ threshold.

In the left panel of Figure 3, we showed $L_{\rm X}$ versus
$\LOIII$. The red squares denote the HBLRs Sy2s and the black
squares for the non-HBLRs Sy2s. The solid line is the best fit
with the fixed slope index of 1.0, $log \LOIII=(-0.91\pm
0.14)+logL_{\rm X}$, with the correlation coefficient (R) of 0.71
and a null probability of $P_{\rm null}<0.001$. The dash line is
the best fit with unfixed slope index, $log \LOIII=(15.18\pm
5.11)+(0.62\pm 0.12)logL_{\rm X}$, R=0.71 and $P_{\rm
null}=0.001$. For two local AGN samples selected by the [O III]
flux and the hard X-ray luminosity, Heckman et al. (2005) found
the relation between the hard X-ray luminosity and the [O III]
luminosity, they found that the mean value of log($L_{\rm
X}/\LOIII$) for Sy2s is 0.57 with a standard deviation of 1.06 dex
(see their Figure 3). Our result is different from theirs. The
reason is that they didn't perform the intrinsic dust extinction
correction for [O III] flux. Recently, Netzer et al. (2006) find
that $ \LOIII /L_{\rm X}$ decrease with $L_{\rm X}$, and give
$log(\LOIII /L_{\rm X})=(15.0\pm 4.0) - (0.38\pm 0.09)log L_{\rm
X}$, which is very similar to our results (see also Shu et al.
2006).

In the right panel of Figure 3, we show $L_{\rm X}/L_{\rm Edd}$
versus $L_{\rm bol}/L_{\rm Edd}$. The solid line is the best fit
with the fixed slope index of 1.0, $log(L_{\rm
bol}/L_{Edd})=(2.63\pm 0.18)+log(L_{\rm X}/L_{\rm Edd})$, R=0.58
and $P_{\rm null}<0.001$. The dash line is the best fit with
unfixed slope index, $log(L_{\rm bol}/L_{\rm Edd}) = (0.95\pm
0.52)+(0.47\pm 0.16)log(L_{\rm X}/L_{\rm Edd}$), R=0.58 and
$P_{\rm null}=0.006$. Considering the relation between $L_{\rm
X}/L_{\rm Edd}$ and $L_{\rm bol}/L_{\rm Edd}$ (the solid line in
the right panel of Figure 3), the $log(L_{\rm X}/L_{\rm Edd})$
threshold of -4 is corresponding to the $log(L_{\rm bol}/L_{\rm
Edd})$ threshold of $-1.37\pm 0.18$.

Following Nicastro et al. (2003), in Figure 4 we show the
distribution of SMBHs  masses (left panel) and the
extinction-corrected $\OIII \lambda$5007 luminosity (right panel)
versus the Eddington ratios. With the larger sample, there is no
obvious boundary in $L_{\rm bol}/L_{\rm Edd}$ (the threshold of
$10^{-1.37\pm 0.18}$) between HBLRs Sy2s and non-HBLRs Sy2s. Below
the Eddington ratio threshold of $10^{-1.37}$, there are only
seven objects, all but one object (NGC513) belong to non-HBLRs
Sy2s. Thus it suggested that at lower Eddington ratio, few HBLRs
Sy2s are found,  which is consistent with the model of Nicastro et
al. (2003).  However, we note that at higher Eddington ratio,
distribution of SMBHs masses and Eddington ratios are nearly the
same for HBLRs Sy2s and non-HBLRs Sy2s (see Figure 4).

\subsection{Other factors on the detectability of
HBLRs in Sy2s: $N_{\rm H}$, [O III]/X-ray luminosity, IRAS color
$f_{60}/f_{25}$ }

Because spectropolarimetry just deals with a few percent of the
light from the object, detectability of HBLRs strongly depends
observationally on a variety of factors, such as the dilution of
the polarized signal by host galaxy light, sensitivity and the
inclination of the torus to the line of sight. It happens that
objects previously thought to be non-HBLR do in fact show broad
polarized lines in deeper observations (e.g., Moran et al 2001).
Here we just use their HBLR/non-HBLR classification from the
sample of Gu \& Huang (2002), which is mostly from Tran (2001;
2003).

As we mentioned in the section of introduction, for Compton-thick
Sy2s ($N_{\rm H}> 10^{24} \rm cm^{-2}$), it is not possible to
estimate precisely the intrinsic X-ray luminosity. In Figure 5,
Compton-thick Sy2s are located above the solid horizon line of
$N_{\rm H} = 10^{24} \rm cm^{-2}$ and Compton-thin Sy2s are below
this line. In the left-hand panel in Figure 5, we show $N_{\rm
H}$ versus $L_{\rm bol}/L_{\rm Edd}$. If just considering the
Compton-thin Sy2s, we find that all three objects below the
threshold of $10^{-1.37}$ are non-HBLRs Sy2s. Although only a few data points,
there is a trend that at smaller Eddington ratio, more
non-HBLRs Sy2s can be found. Above the threshold of $10^{-1.37}$,
there are 10 HBLRs Sy2s and 7 non-HBLRs Sy2s with $L_{\rm
bol}/L_{\rm Edd}$ above the threshold of $10^{-1.37}$. Thus
 at larger Eddington ratio, the detectability of HBLRs
Sy2s and no-HBLRs Sy2s are almost the same (see Figure 5). There are
four unabsorbed (3 non-HBLRs and 1 HBLR) Sy2s below the solid horizon line of $N_{\rm H} =
10^{22} \rm cm^{-2}$ in the left-panel of Figure 5. Two unabsorbed non-HBLRs Sy2s have the
Eddington ratio larger than the threshold of $10^{-1.37}$.

In the right-hand panel in Figure 5, we show $N_{\rm H}$ versus
$\LOIII$. For Compton-thin Sy2s with lower $\OIII$ luminosity
($<10^{41} \ergs$), we find that there are only one HBLRs Sy2s
and 8 non-HBLRs Sy2s, suggesting higher fraction of Compton-thin
non-HBLRs Sy2s. In fact, for the low-luminosity sources ($<10^{41}
\ergs$), there are only two HBLRs Sy2s and 10 non-HBLRs Sy2s,
nearly all of them do not show HBLRs regardless of $N_H$,
suggesting that AGN luminosity plays a major role in the low-luminosity
end. For Compton-thin Sy2s with higher [O III] luminosity
($>10^{41} \ergs$), we find that there are 17 HBLRs Sy2s and 3
non-HBLRs Sy2s, suggesting higher detectability of Compton-thin
HBLRs Sy2s ($\sim$ 85\%). However, for Compton-thick Sy2s with higher [O
III] luminosity ($>10^{41} \ergs$), we found that there are 8
HBLRs Sy2s and 10 non-HBLRs Sy2s, suggesting the detectability
fraction is much smaller than Compton-thin sources. These results are
consistent with the recent work by Shu et al. (2006). Zakamsa et
al. (2003b) conducted spectropolarimetry of 12 type II (obscured)
quasar candidates selected from Sloan Digital Sky Survey (SDSS).
The $\LOIII$ luminosity of these candidates were about
$10^{9}\Lsolar, \sim 10^{42} \ergs$, larger than our adopted value
of $10^{41} \ergs$. Polarizations were detected in all objects and
HBLRs were safely detected in five objects.

It has been showed that HBLRs Sy2s have smaller values of
$f_{60}/f_{25}$, and smaller Balmer decrements as measured by the
ratio of the narrow lines, H$\alpha$/H$\beta$, compared to
non-HBLRs Sy2s (Heisler, Lumsden \& Bailey 1997). The mean value
of $f_{60}/f_{25}$ is $5.08\pm 0.39$ for non-HBLRs Sy2s and
$2.50\pm 0.21$ for HBLRs Sy2s.  Just considering the absorbed Sy2s
($N_{\rm H} > 10^{22} \rm cm^{-2}$), the difference of the
$f_{60}/f_{25}$ distribution is more significant( $5.62\pm 0.67$
for non-HBLRs Sy2s and $2.46\pm 0.23$ for HBLRs Sy2s). And K-S
tests show that the $f_{60}/f_{25}$ distributions for non-HBLRs
Sy2s and HBLRs Sy2s subsamples are drawn from the same parent
population with a possibility of 0.0037 and 0.0031  for the whole
sample and for objects with $N_{\rm H} > 10^{22} \rm cm^{-2}$,
respectively. A plot of $L_{\rm bol}/L_{\rm Edd}$ vs.
$f_{60}/f_{25}$ showed that the correlation would be nice for
Compton-thin Sy2s (R=-0.37, $P_{\rm null}=0.03$). The solid line
is the best fit, $f_{60}/f_{25} = (4.22\pm 0.54)-(1.18\pm
0.52)log(L_{\rm bol}/L_{\rm Edd}$) (see Figure 6). However, this
correlation  becomes weaker (R=-0.32, $P_{\rm null}=0.02$) for the
whole Sy2s sample. Near/mid infrared emission is regarded as being
anisotropic whereas far infrared is isotropic. $f_{60}/f_{25}$ can
be used as an orientation indicator of torus (Heisler et al.
1997). The larger values of $f_{60}/f_{25}$ showed the higher
tours inclination to the line of sight. The median correlation
between $f_{60}/f_{25}$ and $L_{\rm bol}/L_{\rm Edd}$ showed that
the larger Eddington ratio would lead to smaller torus opening
angle (e.g. Wang et al. 2005), and cooler infrared color is
expected (smaller $f_{60}/f_{25}$).

\section{Conclusion}
We calculated the SMBHs masses and the Eddington ratios for a
larger compiled Sy2s sample. For Sy2s with the Eddington ratio
larger than the threshold of $10^{-1.37}$, there is no obvious
discrimination in the Eddington ratios/black hole masses for Sy2s
with and without HBLRs; Sy2s with low luminosity and low Eddington
ratios do not show HBLRs regardless of $N_{\rm H}$, which is
consistent with the prediction of Nicastro et al. (2003). For
high-luminosity Compton-thin Sy2s, we find very higher
detectability of HBLRs Sy2s ($\sim$ 85\%).  However, as the
present sample in this paper is an amalgamation of different
observations with diverse quality of spectropolarimetric data,
varying from object to object determined by the  brightness,
observers, integration time, and a host of other factors, it is
hard to entangle the physical nature of Sy2s with and without
HBLRs. In the future, we need more new data of hard X-ray spectra
with good quality optical spectropolarimetric information.

\section*{ACKNOWLEDGMENTS}
We thank the anonymous referee for her/his suggestions. This work
has been supported by the NSFC (Nos. 10403005, 10473005, 10325313,
10233030 and 10521001) and the Science-Technology Key Foundation
from Education Department of P. R. China (No. 206053). QGU would
like to acknowledge the financial supports from China Scholarship
Council (CSC) and the National Natural Science Foundation of China
under grants 10103001, 10221001 and 10633040. We are grateful to
Dr. Jian-Min Wang for very helpful suggestions.


\newpage
\begin{table*}
\caption{\small The non-HBLRs Sy2s sample. }
\begin{tiny}
\begin{tabular}{lllllllllll}
\hline \hline
name   &$\LOIII$& $\sigma$ & Ref.& $log M_{\rm bh}$ & $log (L_{\rm bol}/L_{\rm Edd})$ & $f_{60}/f_{25}$&$log N_{\rm H}$   & $log L_{\rm X}$ & $log (L_{\rm X}/L_{\rm Edd})$\\
      (1) & (2) &(3) & (4) & (5) & (6)&(7)& (8) & (9)& (10) \\
\hline
Mrk334             &       41.289  &  79     &     4              &            6.51 & 0.22  & 4.04 & 20.64& 43.10& -1.51\\
NGC34              &       42.772  &  105    &     3               &            7.00 & 1.22  & 6.76 & -     & --    & --   \\
IRAS00198-7926     &       42.562  &  239    &     5               &            8.44 & -0.43 & 2.49 & $>24$    & --    & --   \\
Mrk573             &       42.001  &  123$*$ &     1               &            7.28 & 0.16  & 1.59 & -     & --    & --   \\
NGC1144            &       42.251  &  219$*$ &     2               &            8.29 & -0.59 & 8.72 & 20.70& $<43.28$& -3.11\\
Mrk1066            &       42.175  &  105$*$ &     1               &            7.01 & 0.61  & 4.86 & $>24$    & --    & --   \\
NGC1241            &       42.472  &  136$*$ &     2               &            7.46 & 0.46  & 8.29 & -     & --    & --   \\
NGC1320            &       40.959  &  116$*$ &     1               &            7.18 & -0.78 & 2.19 & -     & --    & --   \\
NGC1358            &       40.783  &  173$*$ &     1               &            7.88 & -1.65 & 3.05 & -     & --    & --   \\
NGC1386            &       40.586  &  120$*$ &     1               &            7.24 & -1.21 & 4.03 & $>25$    & --    & --   \\
IRAS03362-1642     &       41.529  &  --     &     -               &            --   & --    & 2.13 & -     & --    & --   \\
IRAS04103-2838     &       -       &  --     &     -               &            --   & --    & 3.39 & -     & --    & --   \\
IRAS04210+0400     &       42.328  &  127    &     6               &            7.34 & 0.44  & 2.38 & -     & --    & --   \\
IRAS04229-2528     &       41.903  &  --     &     -               &            --   & --    & 3.69 & -     & --    & --   \\
IRAS04259-0440     &       40.44   &  --     &     -               &            --   & --    & 2.80 & -     & --    & --   \\
NGC1667            &       41.921  &  173$*$ &     1               &            7.88 & -0.51 & 8.91 & $>24$    & --    & --   \\
NGC1685            &       42.582  &  --     &     -               &            --   & --    & 4.48 & -     & --    & --   \\
ESO428-G014        &       41.937  &  127    &     7               &            7.34 & 0.05  & 2.49 & $>25$    & --    & --   \\
IRAS08277-0242     &       -       &  --     &     -               &            --   & --    & 3.45 & -     & --    & --   \\
NGC3079            &       40.427  &  150$*$ &     8               &            7.63 & -1.76 & 14.01 & 22.204& 40.25& -5.48\\
NGC3281            &       40.998  &  123    &     3               &            7.28 & -0.84 & 2.65 & 23.903& 42.80& -2.58\\
IRAS10340+0609     &       -       &  --     &     -               &            --   & --    & 1.56 & -     & --    & --   \\
NGC3362            &       41.269  &  92$*$  &     1               &            6.77 & -0.06 & -     & -     & --    & --   \\
UGC6100            &       42.18   &  156$*$ &     1               &            7.70 & -0.07 & 2.84 & -     & --    & --   \\
NGC3660            &       40.914  &  --     &     -               &            --   & --    & 7.78  & 20.255& 41.8  & --   \\
NGC3982            &       40.019  &  62$*$  &     1               &            6.09 & -0.62 & 8.26 & -     & --    & --   \\
NGC4117            &       -       &  95$*$  &     1               &            6.83 & --    & -     & -     & --    & --   \\
NGC4501            &       39.804  &  171$*$ &     8               &            7.86 & -2.61 & 5.95 & -     & --    & --   \\
NGC4941            &       40.894  &  80     &     3               &            6.53 & -0.19 & 4.10 & 23.653& 40.82 & -3.81\\
NGC5135            &       42.311  &  128$*$ &     2               &            7.35 & 0.40  & 6.73 & $>24$    & --    & --   \\
NGC5194            &       40.168  &  102$*$ &     1               &            6.95 & -1.34 & 6.59 & 23.699& 39.96& -5.10\\
NGC5256            &       41.825  &  100    &     3               &            6.92 & 0.35  & 6.36 & $>24$    & --    & --   \\
NGC5283            &       40.836  &  148$*$ &     1               &            7.60 & -1.32 & -     & -     & --    & --   \\
Mrk1361            &       42.242  &  --     &           -         &            --   & --    & 3.91 & -     & --    & --   \\
IRAS13452-4155     &       42.098  &  80     &     9               &            6.51 & 1.03  & 2.27 & -     & --    & --   \\
NGC5643            &       41.225  &  77     &     3               &            6.44 & 0.23  & 5.03 & $>25$    & --    & --   \\
NGC5695            &       40.501  &  144$*$ &     1               &            7.56 & -1.61 & 4.39 & -     & --    & --   \\
NGC5728            &       42.083  &  209$*$ &     8               &            8.21 & -0.68 & 10.19 & -     & --    & --   \\
IRAS19254-7245     &       42.731  &  --     &     -               &            --   & --    & 3.96 & $>24$    & --    & --   \\
NGC6890            &       40.808  &  78     &     3               &            6.48 & -0.23 & 4.92 & -     & --    & --   \\
IRAS20210+1121     &       43.124  &  --     &     -               &            --   & --    & 2.42 & $>25$    & --    & --   \\
NGC7130            &       42.477  &  141$*$ &     2               &            7.52 & 0.40  & 7.76 & $>24 $   & --    & --   \\
NGC7172            &       39.767  &  154$*$ &     2               &            7.67 & -2.46 & 6.37 & 22.934& 42.50& -3.28\\
NGC7496            &       40.211  &  76$*$  &     2               &            6.44 & -0.79 & 5.54 & 22.699& 41.65& -2.89\\
IRAS23128-5919     &       -       &  --     &     -               &            --   & --    & 6.79 & -     & --    & --   \\
IC5298             &       42.078  &  --     &     -               &            --   & --    & 4.65 & -     & --    & --   \\
NGC7582            &       41.357  &  121$*$ &     2               &            7.25 & -0.45 & 6.90 & 23.079& 42.14& -3.21\\
NGC7590            &       39.949  &  93$*$  &     2               &            6.79 & -1.40 & 9.07 & 20.964& 40.80& -4.10\\
NGC7672            &       -       &  98$*$  &     1               &            6.88 & --    & 2.96 & -     & --    & --   \\
\hline

\end{tabular}
{ \noindent \vglue 0.5cm {\sc Note}: Col. (1) the galaxy name;
Col. (2) the extinction-corrected $\OIII \lambda$5007 luminosity
in units of $\Lsolar$; Col. (3) the stellar velocity dispersion in
units of \kms; Col. (4) reference of $\sigma_{*}$; Col. (5) SMBHs
masses in units of $\Msolar$; Col. (6) the Eddington ratio; Col.
(7) IRAS color $f_{60}/f_{25}$; Col. (8) gaseous absorbing column
density ($N_{\rm H}$) in units of $\rm cm^{-2}$; Col. (9) the
absorption-corrected hard X-ray (2-10 keV) luminosity in units of
$\ergs$ for Compton-thin Sy2s; Col. (10) the ratio of $L_{\rm X}$
to the Eddington luminosity. $~~*$ (in Col. 3): $\sigma_{*}$ from
Ca II Triplets, and the rest are from FWHM of [O III]$\lambda$
5007.

{\sc References.}:
(1) Nelson \& Whittle 1995; (2) Garcia-Rissmann
2005; (3) Whittle 1992; (4) Wilson \& Nath 1990; (5) Lipariet et
al. 1991; (6) Gelderman \& Whittle 1994; (7) Wilson \& Baldwin
1989; (8) McElroy 1995; (9) Young et al. 1996. }
\end{tiny}
\end{table*}

\begin{table*}
\caption{\small The HBLRs Sy2s sample. }
\begin{tiny}
\begin{tabular}{lllllllllll} \hline \hline
name   &$\LOIII$& $\sigma$ & Ref.& $log M_{\rm bh}$ & $log (L_{\rm bol}/L_{\rm Edd})$ & $f_{60}/f_{25}$&$log N_{\rm H}$   & $log L_{\rm X}$ & $log (L_{\rm X}/L_{\rm Edd})$\\
      (1) & (2) &(3) & (4) & (5) & (6)&(7)& (8) & (9)& (10) \\
\hline
Mrk348             &       41.912  &  118$*$ &     1               &            7.21 & 0.15  & 1.42 & 23.04& 43.01& -2.30\\
IRAS00521-7054     &       42.743  &  --     &     -               &            --   & --    & 1.27  & -     & --    & --   \\
NGC424             &       41.462  &  163    &     3               &            7.78 & -0.87 & 1.18 & 24.18& --    & --   \\
NGC513             &       40.597  &  152$*$ &     1               &            7.65 & -1.61 & 6.33 & -     & --    & --   \\
NGC591             &       41.954  &  95$*$  &     1               &            6.83 & 0.57  & 4.33 & -     & --    & --   \\
IRAS01475-0740     &       41.689  &  --     &     -               &            --   & --    & 1.10 & -     & --    & --   \\
NGC788             &       40.99   &  140$*$ &     1               &            7.51 & -1.07 & 0.99 & -     & --    & --   \\
NGC1068            &       42.645  &  128$*$ &     1               &            7.35 & 0.74  & 2.16  & $>25$    & --    & --   \\
IRAS02581-1136     &       41.536  &  --     &     -               &            --   & --    & 1.07 & -     & --    & --   \\
IRAS04385-0828     &       40.117  &  --     &     -               &            --   & --    & 1.77 & -     & --    & --   \\
IRAS05189-2524     &       42.459  &  --     &     -               &            --   & --    & 3.99 & 22.69 & 43.33& --   \\
Mrk3               &       43.221  &  269$*$ &     1               &            8.65 & 0.02  & 1.30   & 24.04& --    & --   \\
NGC2273            &       41.449  &  124$*$ &     1               &            7.30 & -0.40 & 4.88 & $>25$    & --    & --   \\
Mrk1210            &       42.195  &  77$*$  &     2               &            6.46 & 1.18  & 0.98 & $>24$    & --    & --   \\
NGC3081            &       41.331  &  129$*$ &     2               &            7.36 & -0.59 & -     & 23.82 & 41.83& -3.63\\
IRAS11058-1131     &       42.316  &  --     &     -               &            --   & --    & 2.39 & $>24$    & --    & --   \\
Was49b             &       42.412  &  --     &     -               &            --   & --    & 1.45 & 22.799& 42.97 & --   \\
NGC4388            &       41.684  &  119$*$ &     1               &            7.22 & -0.10 & 2.94 & 23.623& 42.74& -2.58\\
NGC4507            &       41.569  &  146$*$ &     2               &            7.58 & -0.57 & 3.95 & 23.462& 43.22& -2.46\\
IC3639             &       42.113  &  95$*$  &     2               &            6.83 & 0.73  & 2.95 & $>25$    & --    & --   \\
MCG-3-34-64        &       42.322  &  --     &     -               &            --   & --    & 2.06 & 23.881& 42.53& --   \\
NGC5252            &       41.963  &  190$*$ &     1               &            8.04 & -0.63 & -     & 22.633& 43.06& -3.08\\
NGC5347            &       40.446  &  93$*$  &     1               &            6.79 & -0.90 & 1.25 & $>24$    & --    & --   \\
Mrk463E            &       42.785  &  173    &     3               &            7.88 & 0.35  & 1.51 & 23.204& 42.65& -3.33\\
Circinus           &       40.474  &  75$*$  &     1               &            6.42 & -0.50 & 3.64  & $>24$    & --    & --   \\
NGC5506            &       41.608  &  86     &     4               &            6.65 & 0.40  & 2.01 & 22.531& 42.86& -1.89\\
Mrk477             &       43.543  &  117    &     3               &            7.20 & 1.79  & 2.51 & $>24$    & --    & --   \\
ESO273-IG04        &       42.372  &  --     &     -               &            --   & --    & 2.77 & -     & --    & --   \\
NGC5929            &       40.926  &  121$*$ &     1               &            7.25 & -0.88 & 5.64 & 20.763& 42.09& -3.27\\
NGC5995            &       42.904  &  --     &     -               &            --   & --    & 2.82 & 21.934& 43.43& --   \\
IRAS15480-0344     &       42.946  &  --     &     -               &            --   & --    & 1.47 & -     & --    & --   \\
IRAS17345+1124     &       42.956  &  --     &     -               &            --   & --    & 2.47 & -     & --    & --   \\
NGC6552            &       42.122  &  --     &     -               &            --   & --    & 3.64 & 23.778& 42.46& --   \\
IRAS20460+1925     &       42.838  &  --     &     -               &            --   & --    & 1.67 & 22.398& 44.08& --   \\
IC5063             &       41.919  &  160$*$ &     1               &            7.74 & -0.38 & 1.53 & 23.38 & 42.85& -2.99\\
IRAS22017+0319     &       42.497  &  --     &     -               &            --   & --    & 1.61 & 21.301& 43.41 & --   \\
NGC7212            &       42.636  &  137$*$ &     1               &            7.47 & 0.61  & 4.08 & 23.653& 42.36& -3.21\\
MCG-3-58-7         &       43.916  &  --     &     -               &            --   & --    & 2.77 & -     & --    & --   \\
IRAS23060+0505     &       41.692  &  --     &     -               &            --   & --    & 2.70   & 22.924& 44.16& --   \\
NGC7674            &       42.495  &  144$*$ &     1               &            7.56 & 0.38  & 2.81 & $>25$    & --    & --   \\
NGC7682            &       41.718  &  123$*$ &     1               &            7.28 & -0.12 & 2.03 & -     & --    & --   \\
\hline

\end{tabular}
{ \noindent \vglue 0.5cm {\sc Note}: $~~*$ (in Col. 3):
Col. (1)
the galaxy name; Col. (2) the extinction-corrected $\OIII
\lambda$5007 luminosity in units of $\Lsolar$; Col. (3) the
stellar velocity dispersion in units of \kms; Col. (4) reference
of $\sigma_{*}$; Col. (5) SMBHs masses in units of $\Msolar$; Col.
(6) the Eddington ratio; Col. (7) IRAS color $f_{60}/f_{25}$; Col.
(8) gaseous absorbing column density ($N_{\rm H}$) in units of
$\rm cm^{-2}$; Col. (9) the absorption-corrected hard X-ray (2-10
keV) luminosity in units of $\ergs$ for Compton-thin Sy2s; Col.
(10) the ratio of $L_{\rm X}$ to the Eddington luminosity.
$\sigma_{*}$ from Ca II Triplets, and the rest are from FWHM of [O
III]$\lambda$ 5007.

{\sc References.}:
(1) Nelson \& Whittle 1995; (2) Garcia-Rissmann
2005; (3) Whittle 1992; (4) Wilson \& Nath 1990; (5) Lipariet et
al. 1991; (6) Gelderman \& Whittle 1994; (7) Wilson \& Baldwin
1989; (8) McElroy 1995; (9) Young et al. 1996. }
\end{tiny}

\end{table*}

\begin{table*}
\caption{The distributions of SMBHs masses and the Eddington
ratios for HBLRs Sy2s and non-HBLRs Sy2s.}
\begin{small}
\begin{center}
\begin{tabular}{llllllllllllll}
\hline \hline
Type & log($M_{BH}/\Msolar$) & $SD_{1}$ & N1 & P1 & log($L_{bol}/L_{Edd}$) & $SD_{2}$ & N2 & P2\\
(1)& (2) & (3) & (4) & (5) & (6) & (7)& (8) & (9)  \\
\hline
non-HBLRs Sy2s &  7.22$\pm$ 0.10     &    0.57 & 36 &  64.76\% & -0.47$\pm$ 0.16     &    0.93 & 34  & 48.50\%\\
HBLRs Sy2s     &  7.33$\pm$ 0.10     &    0.51 & 24 &          & -0.07$\pm$ 0.16     &    0.79 & 24  & \\

A:non-HBLRs Sy2s &  7.35$\pm$ 0.19     &    0.57 & 9 & 82.41\% & -0.01$\pm$ 0.20     &    0.59 & 9 & 43.46\% \\
A:HBLRs Sy2s     &  7.23$\pm$ 0.21     &    0.67 & 10 &        &  0.22$\pm$ 0.29     &    0.90 & 10 & \\

B:non-HBLRs Sy2s &  7.13$\pm$ 0.19     &    0.60 & 10 & 40.05\% & -0.96$\pm$ 0.25     &    0.79 & 10 & 16.70\% \\
B:HBLRs Sy2s     &  7.44$\pm$ 0.13     &    0.40 & 10 &         &-0.16$\pm$ 0.16     &    0.52 & 10  & \\

C:non-HBLRs Sy2s &  7.33$\pm$ 0.11     &    0.53 & 25 & 93.86\% & -0.78$\pm$ 0.19     &    0.91 & 23 & 16.53\% \\
C:HBLRs Sy2s     &  7.33$\pm$ 0.12     &    0.52 & 20 &         &-0.17$\pm$ 0.16     &    0.72 & 20  & \\

\hline \hline
\end{tabular}
\parbox{6.5in}
{\baselineskip 9pt \noindent \vglue 0.5cm {\sc Note}:Col.1: type;
Col.2: log of SMBHs masses in units of \Msolar; Col.3: the
standard deviation of log of SMBHs masses; Col.4: Number; Col.5:
possibility from the same parent population of the SMBHs mass
distribution; Col.6: log of the Eddington ratio; Col.7: the
standard deviation of log of the Eddington ratio; Col.8: Number.
Col. 9: possibility from the same parent population of the
Eddington ratio. A: objects with $N_{\rm H}\geq 10^{24} \rm cm
^{-2}$ and the available Eddington ratios, B: objects with $N_{\rm
H} < 10^{24} \rm cm ^{-2}$ and the available Eddington ratios, C:
objects with the direct $\sigma_*$ measurements.}

\end{center}
\end{small}
\end{table*}

\newpage

\begin{figure*}
\begin{center}
\includegraphics{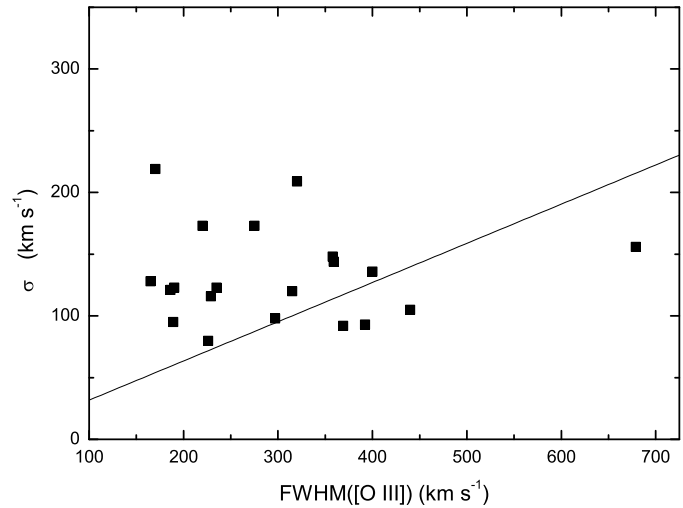}
\caption{The [O III] FWHM versus $\sigma_{*}$ for the common
objects in our sample and the sample of Zhang \& Wang (2006). The
solid line showed $\sigma_{*}=FWHM_{\rm [O III]}/2.35/1.34$. }
\end{center}
\end{figure*}

\begin{figure*}
\begin{center}
\includegraphics[width=8cm,height=12cm]{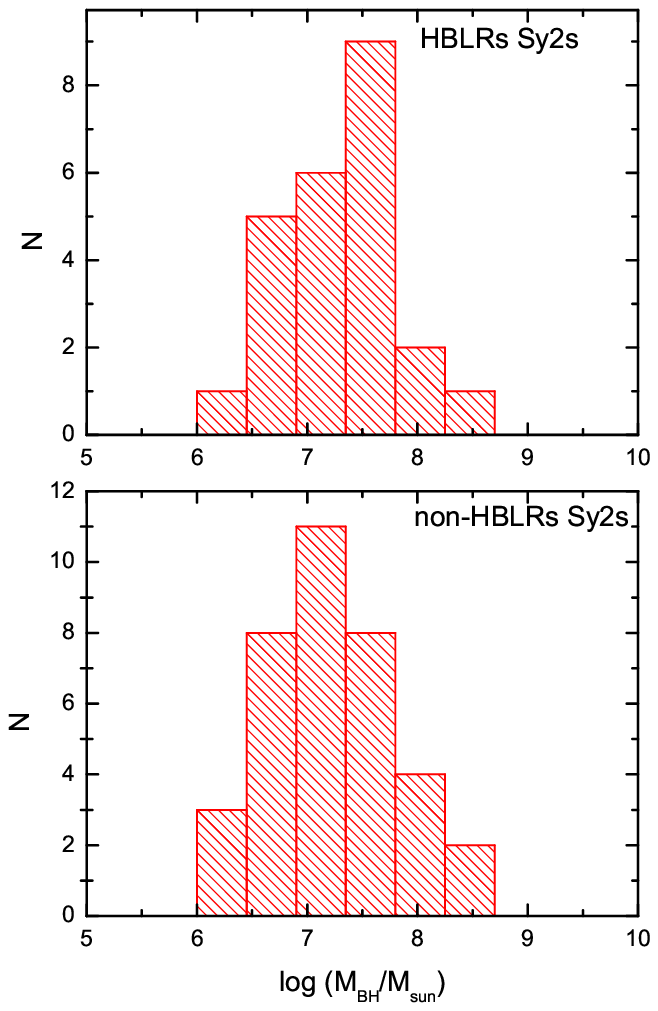}
\includegraphics[width=8cm,height=12cm]{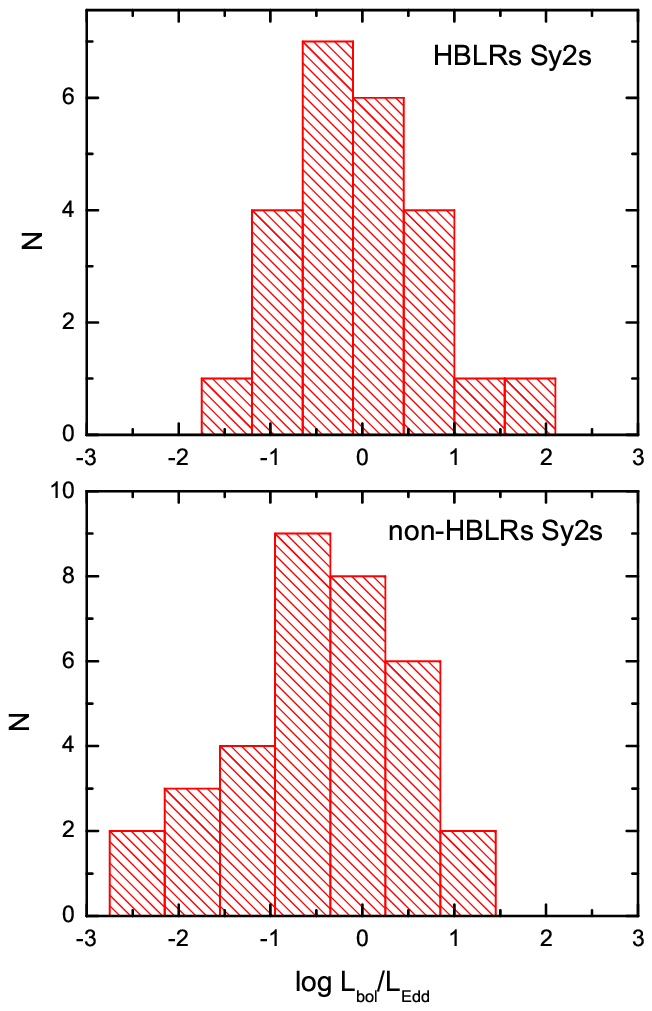}
\caption{Left-hand side: the distribution of SMBHs masses for
HBLRs Sy2s and non-HBLRs Sy2s. Right-hand side: the distribution
of the Eddington ratios for HBLRs Sy2s and non-HBLRs Sy2s.}
\end{center}
\end{figure*}

\begin{figure*}
\begin{center}
\includegraphics[width=8cm,height=6cm]{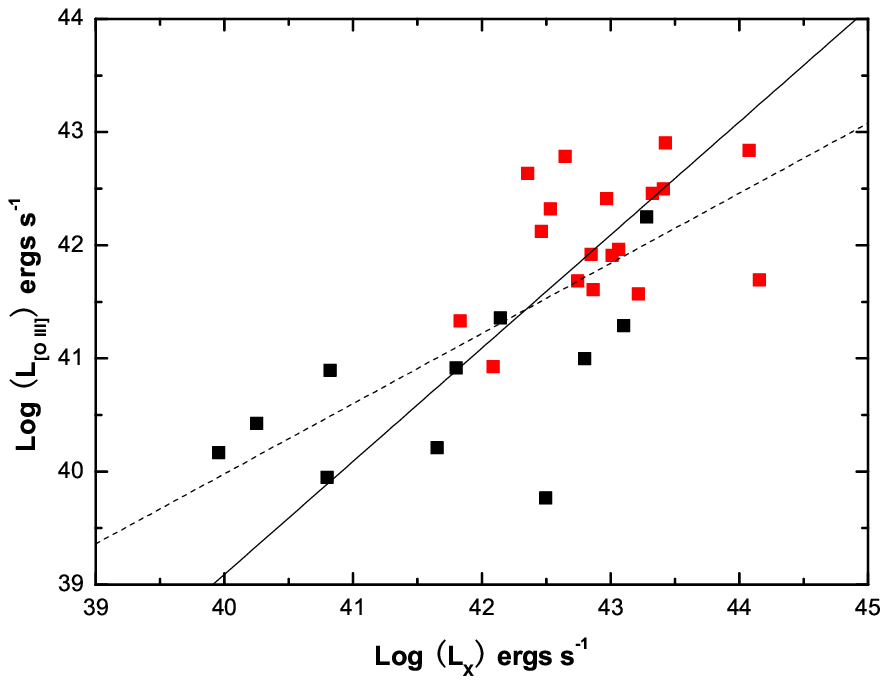}
\includegraphics[width=8cm,height=6cm]{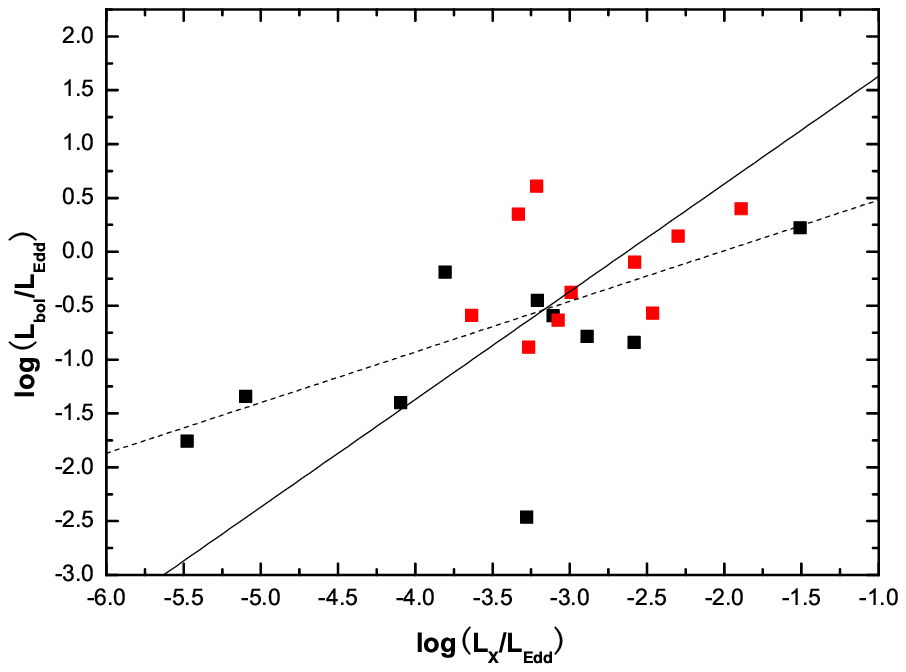}
\caption{Left-hand side: $L_{\rm X}$ versus $\LOIII $; Right-hand
side: $L_{\rm X}/L_{\rm Edd}$ versus $L_{\rm bol}/L_{\rm Edd}$.
The solid line is the best fit with the fixed slope index of one
and the dash line is the best fit with unfixed slope index. The
red squares denote HBLRs Sy2s and the black squares denote
non-HBLRs Sy2s.}
\end{center}
\end{figure*}

\begin{figure*}
\begin{center}
\includegraphics[width=8cm,height=6cm]{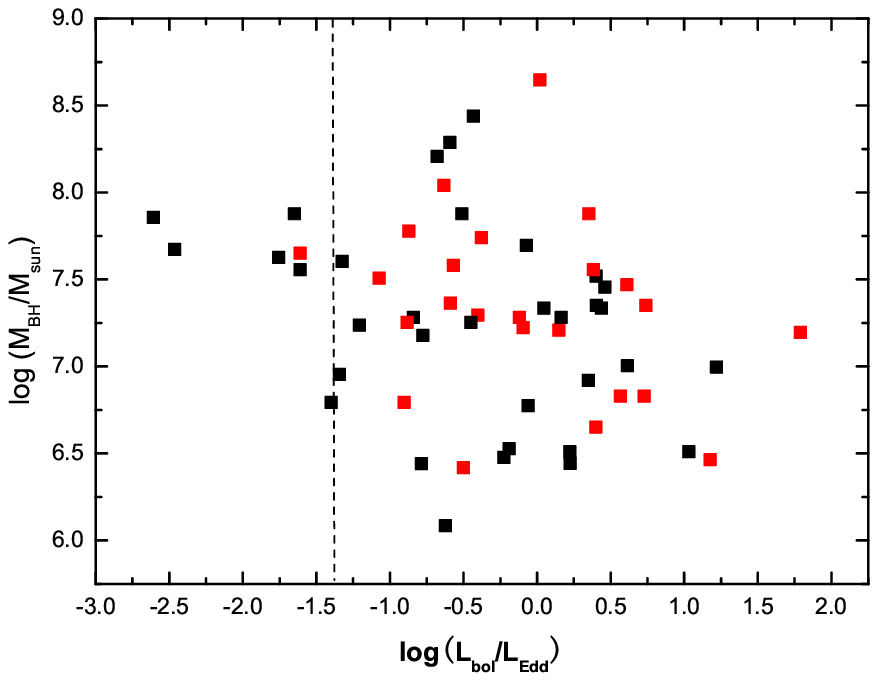}
\includegraphics[width=8cm,height=6cm]{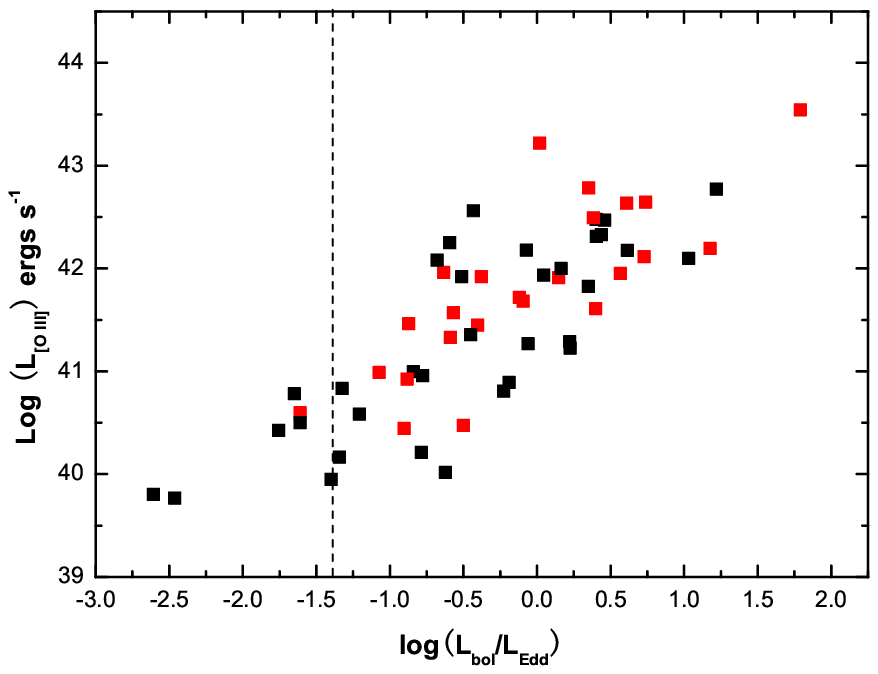}
\caption{Left-hand side: the central SMBHs masses versus the
Eddington ratios. Right-hand side: the [O III] $\lambda$ 5007
luminosity versus the Eddington ratios. The dash line showed the
$log (L_{\rm bol}/L_{\rm Edd})$ threshold of -1.37. The symbols
are same to that in Fig. 3.}
\end{center}
\end{figure*}

\begin{figure*}
\begin{center}
\includegraphics[width=8cm,height=6cm]{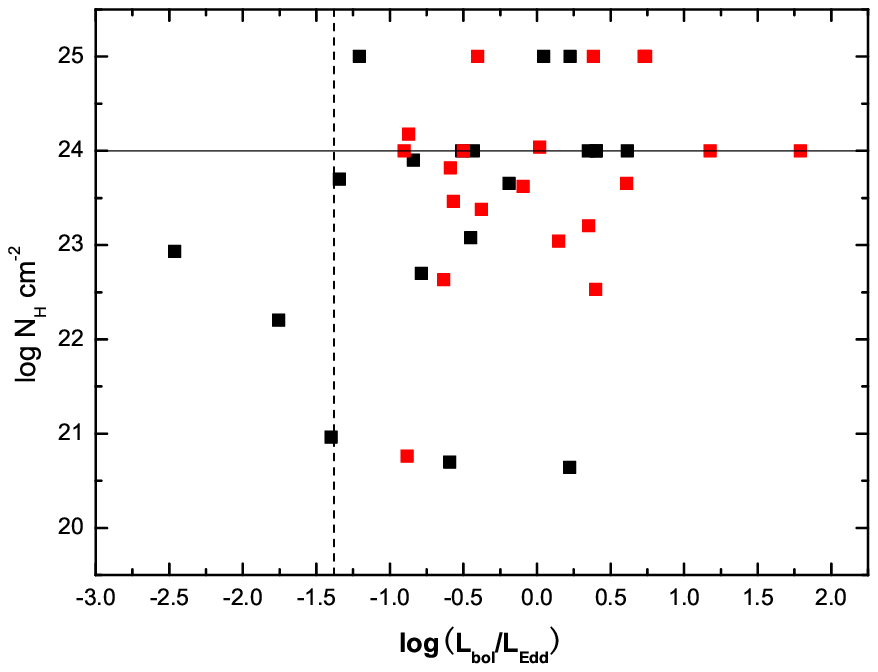}
\includegraphics[width=8cm,height=6cm]{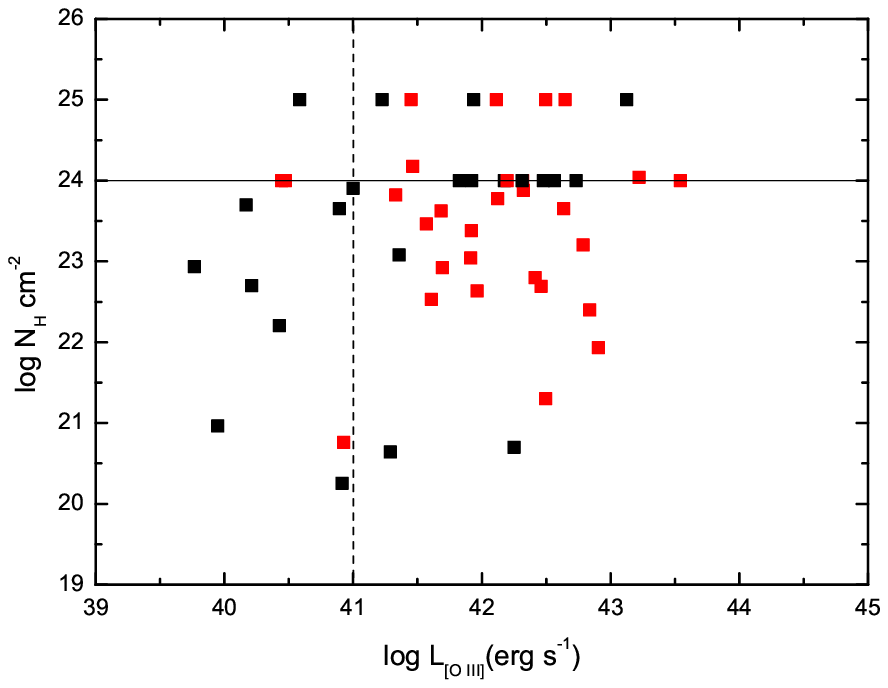}
\caption{Left-hand side: $N_{\rm H}$ versus the Eddington ratios.
Right-hand side: $N_{\rm H}$ versus the [O III] $\lambda$ 5007
luminosity. The dash line showed the $log (L_{bol}/L_{\rm Edd})$
threshold of -1.37 (left panel) and the $\LOIII$ threshold of
$10^{41} erg s^{-1}$ (right panel). The horizon solid line is the
line of $N_{\rm H}=10^{24} \rm \rm cm ^{-2}$. The symbols are same
to that in Fig. 3.}
\end{center}
\end{figure*}

\begin{figure*}
\begin{center}
\includegraphics[width=8cm,height=6cm]{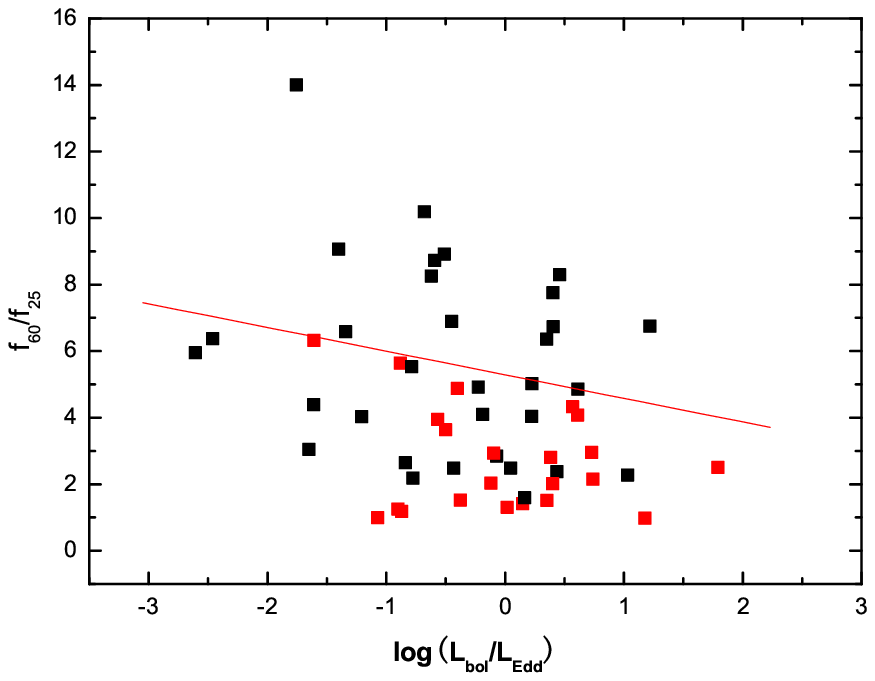}
\caption{$f_{60}/f_{25}$ versus $L_{\rm bol}/L_{\rm Edd}$ for
Compton-thin Sy2s ($N_{\rm H} < 10^{24} \rm \rm cm ^{-2}$). The
symbols are same to that in Fig. 3.}
\end{center}
\end{figure*}


\begin{thebibliography}{}
\bibitem[]{} Antonucci, R., 1993, ARA\&A, 31, 473
\bibitem[]{} Antonucci, R.,  Miller J.S., 1985, ApJ, 297, 621
\bibitem[]{} Baskin, A., Laro, A., 2005, MNRAS, 358, 1043
\bibitem[]{} Bassani, L., et al. 1999, ApJS, 121, 473
\bibitem[]{} Bian, W., Gu, Q., Zhao, Y., et al.,  2006, MNRAS, 372,
876
\bibitem[]{} Bian, W., Zhao, Y.,  2002, A\&A, 395, 465
\bibitem[]{} Bian, W., Zhao, Y.,  2004, MNRAS, 347, 607

\bibitem[]{} Gallo, L. C., et al., 2006, MNRAS, 365, 688
\bibitem[]{} Garcia-Rissmann, A., Vega, L. R., Asari, N. V., et al., 2005, MNRAS, 359, 765
\bibitem[]{} Gelderman, R., Whittle M., 1994, ApJS, 91, 491
\bibitem[]{} Greene, J. E., Ho, L. C., 2005, ApJ, 627, 721
\bibitem[]{} Greene, J. E., Ho, L. C., 2006a, ApJL, 641, L21
\bibitem[]{} Greene, J. E., Ho, L. C., 2006b, ApJ, 641, 117
\bibitem[]{} Gu, M, Cao, X., Jiang D., 2001, MNRAS, 327, 1111
\bibitem[]{} Gu, Q., Huang, J., 2002, ApJ, 579, 205
\bibitem[]{} Heckman, T. M., Kauffmann, G., Brinchmann, J., Charlot S.,
Tremonti C., White, S. D. M., 2004, ApJ, 613, 109
\bibitem[]{} Heckman, T. M., Ptak, A., Hornschemeier, A., Kauffmann, G., 2005, ApJ, 634, 161
\bibitem[]{} Heisler, C. A., Lumsden, S. L.,  Bailey, J. A. 1997, Nature, 385, 700
\bibitem[]{} Kaspi, S., Maoz, D., Netzer, H., Peterson, B.M.,
Vestergaard, M., \& Jannuzi, B.T. 2005, ApJ, 629, 61
\bibitem[]{} Kaspi, S., Smith, P.S., Netzer, H., Maoz, D.,
Jannuzi, B.T., Giveon, U., 2000, ApJ, 533, 631
\bibitem[]{} Laor, A. 2003, ApJ, 590, 86
\bibitem[]{} Lipari, S., Bonatto, C., Pastoriza, M.G., 1991, MNRAS, 253, 19
\bibitem[]{} Lumsden, S.L., Heisler, C.A., Bailey, J.A., Hough, J.H., Young,
S., 2001, MNRAS, 327, 459
\bibitem[]{} Lumsden, S., Alexander, D.,  Hough, J., 2004, MNRAS, 348, 1451
\bibitem[]{} Maiolino, R., et al. 1998, A\&A, 338, 781
\bibitem[]{} McElroy, D. B., 1995, ApJS, 100, 105
\bibitem[]{} McLure, R. J., Jarvis, M. J., 2002, MNRAS, 337, 109
\bibitem[]{} Moran, E.C., Barth, A.J., Kay, L.E., Filippenko, A.V., 2000, ApJ,
540, L73
\bibitem[]{} Moran, E. C., Kay, L. E., Davis, M., Filippenko, A. V., \&
Barth, A. 2001, ApJ, 556, L75
\bibitem[]{} Nelson, C.H., Whittle, M., 1995, ApJS, 99, 67
\bibitem[]{} Nelson, C. H., 2001, ApJ, 544, L91
\bibitem[]{} Netzer, H.  et al. 2006, A\& A, in press, astro-ph/0603712
\bibitem[]{} Nicastro, F., 2000, ApJ, 530, L65
\bibitem[]{} Nicastro, F., Martocchia A., Matt G., 2003, ApJ, 589, L13
\bibitem[]{} Panessa, F. \& Bassani, L., 2002, A\&A, 394, 435
\bibitem[]{} Rees, M. J. 1984, ARA\&A, 22, 471
\bibitem[]{} Shu, et al., 2006, ApJ, in press, astro-ph/0603338
\bibitem[]{} Tran, H. D., 1995, ApJ, 440, 565
\bibitem[]{} Tran, H. D., 2001, ApJ, 554, L19
\bibitem[]{} Tran, H. D., 2003, ApJ, 583, 632
\bibitem[]{} Tremaine, S., et al., 2002, Ap J, 574, 740
\bibitem[]{} Urry, C. M., Padovani, P., 1995, PASP, 107, 803
\bibitem[]{} Veilleux, S., Goodrich, R. W., Hill, G. J., 1997, ApJ, 477, 631
\bibitem[]{} Vestergaard, M., 2002, ApJ, 571, 733
\bibitem[]{} Wang, J. M., Zhang E. P., Luo B., 2005, ApJ, 627, L5
\bibitem[]{} Wang, T. G., Lu, Y. J., 2001, A\&A, 377, 52
\bibitem[]{} Whittle, M. 1992, ApJS, 79, 49
\bibitem[]{} Wilson, A.S., Nath, B., 1990, ApJS, 74, 731
\bibitem[]{} Wilson, A.S., Baldwin, J.A., 1989, AJ, 98, 205
\bibitem[]{} Wu, X. B., et al., 2004, A\& A, 424, 793
\bibitem[]{} Young, S., et al., 1996, MNRAS, 281, 1206
\bibitem[]{} Zakamsa, N. L., et al., 2003a, AJ, 126, 2125
\bibitem[]{} Zakamsa, N. L., et al., 2003b, AJ, 129, 1212
\bibitem[]{} Zhang, E. P., Wang, J. M., 2006, ApJ, in press, astro-ph/0606103



\end{thebibliography}
\end{document}